\documentclass[a4paper]{article}
\usepackage{amsfonts}
\usepackage{amsmath}
\author{Andrzej M. Frydryszak\\
Institute of Theoretical Physics, University of Wroclaw,\\
pl. Borna 9,50-205~Wroclaw, Poland\\
{\tt amfry@ift.uni.wroc.pl}}
\title{\sc \bf Qubits, superqubits and squbits}
\begin{document}
\maketitle

\begin{abstract}
We analyze recently proposed formalisms which use nilpotent variables to describe 
and/or generalize qubits and notion of entanglement. There are two types of them 
distinguished by the commutativity and or anti-commutativity of basics variables.
While nilpotent commuting variables suit the new description of the qubits and entanglement, 
the anticommuting do not, but they can be used to describe generalized objects - superqubits. 
A squbit, in the present context, is a version of superqubit introduced to cure some 
of problematic properties of the superqubit.   
\end{abstract}

\section{Introduction}
New algebraic approaches to study qubit systems and their 
generalizations in the context of quantum entanglement have appeared recently. 
Depending on the basics 
assumption of the variables realizing the two-levelness of the quantum system, 
the final outcome is different. The idea to describe the two level system by specially
chosen variables is around for some time (see account and references in \cite{amf-ncm})
The basics assumption is realized by the condition
\begin{equation}
1=(\Gamma)^0, \quad\quad \Gamma, \quad\quad (\Gamma)^2=0,
\end{equation}
which yelds the use of nilpotents with vanishing square 
(so called nilpotents of order one). 
Obviously one can consider higher order nilpotents to describe multi-level system e.g.
qutrits \cite{mandi}. Realization of nilpotency is very natural in the Grassmann algebra
and variables of this type are used in formulation of supersymmetry since the late seventies of 
the Twentieth Century. Usually such variables are denoted by $\theta$. Another possibility of 
realization of nilpotency, despite the graded commutative Grassmannian algebra, 
gives the commutative algebra with nilpotents: $\eta$, $\eta^2=0$. This option was used by
physicists a bit later \cite{zieg, palu, mandi, amf-cnm, amf-nqm}, firstly within the 
statistical physics to realize the Pauli exclusion principle for spinless objects like fluxes.
Recenly there was developed classical \cite{amf-cnm} and quantum \cite{mandi, amf-nqm} 
formalism based 
on such variables to describe multiqubit systems. This formalism in natural way 
allows to study
multiqubit entanglement and structure of state spaces. 

It turns out that anticommuting $\theta$ variables and the superspaces are not well suited
to study the multiqubit systems and the questions of entanglement. I particular, 
natural realizations of supersymmetry concern rather different systems then qubits, moreover
the conventional Berezinian/superdeterminant is not good for characterization of entanglement. 
It is worth noting that the first time the supersymmetrization of the qubit was proposed by Hruby \cite{hruby},
that was done within the supersymmetric field theoretical model where the qubit field has its anticommuting partner
antiqubit. However there one uses rather field theoretical generalization of qubit to QFT-qubit. 
Here we are focused on nonrelativistic quantum mechanical objects.
\begin{center}
\begin{table}[h]
\caption{\label{comp}Comparison of the $\theta$ and $\eta$ description of two qubit system.}
\centering
\begin{tabular}{ccc}
\vspace*{2mm}\\
\hline
Anticommutative &\phantom{mmmmmmmmm}& Commutative\\
\hline\\
$g(\theta_1, \theta_2)=$ && $f(\eta_1, \eta_2)=$\\
$g_0+\theta_1 g_1+\theta_2 g_2+\theta_1\theta_2 g_{12} $
&& $f_0+\eta_1f_1+\eta_2f_2+\eta_1\eta_2 f_{12} $\\[3mm]
$
Ber g=\left(
\begin{array}{cc}
g_0 & g_1\\
g_2 & g_{12}
\end{array}
\right)
$
&&
$
det f=\left(
\begin{array}{cc}
f_0 & f_1\\
f_2 & f_{12}
\end{array}
\right)
$
\\[3mm]
$=(g_0-g_1g_{12}^{-1}g_2)g_{12}^{-1}$ && $=f_0f_{12}-f_1f_2$
\end{tabular}
\end{table}
\end{center}
Comparing formulas in Tab. \ref{comp} let us observe that expansion coefficients $g_1$ and $g_2$ of the superfunction
$g$ have to be odd elements of a graded algebra and cannot be reduced to the complex 
or real numbers. For $\eta$-function one can take all coefficients $f_i$, $f_{12}$
complex or real valued. This point is essential from the physical point of view. Namely,
when treating $g(\theta_i)$ and $f(\eta_i)$ as a generalized wave functions and developing
generalization of the Schr\"{o}dinger realization for $\theta$ variables one gets Grassmann 
algebra valued scalar product \cite{deW, amf-gns, amf-ghg} and graded algebra valued 
"probability amplitudes". Therefore some additional mapping is needed
to get numerical information from such a Grassmann element. Usualy the so called {\em body mapping} 
\cite{deW} is used, other posibilities gives Banach-Grassmann algebra defined by Rogers \cite{rogers, rogers-b}. 
The $\eta$-functions do not yield such a complications. Such approaches were already 
considered in the context of supersymmetric quantum mechanics in super-Hilbert spaces 
cf. Refs. \cite{nico, deW, amf-gns, amf-ghg}. Using super-Hilbert space approach one can consider
e.g. coherent fermionic states which have properties analogous to the conventional coherent states, 
but are not physical in the usual sense.  
The supermatrix and matrix appearing in the Table \ref{comp} can be thought 
as a matrix of second order $\theta$-superderivative and $\eta$-derivative. Again, in 
the case of the $\theta$-function the properties of the Berezinian prevent us from getting 
information about factorization properties of the $g(\theta_1, \theta_2)$, while conventional
determinant for the $f(\eta_1, \eta_2)$ does the job.  In supersymmetric theories, but not only, 
the anticommuting variables are used to describe fermionic degrees of freedom. As was observed 
by Kitayev in the context of quanum information theory multi-fermion states are more nonlocal
then bosonoc ones, therefore we can expect that objects like superqubit may produce more nonlocality.

One can expect that superspaces related to graded commutative algebra might be not right arena to describe 
qubits, but as it was shown in Ref. \cite{duff_1} specific supersymmetric generalization of
qubit. Moreover for the superqubit systems the notion of entanglement has to be modified
due to some algebraic artefacts. 
As it was noted above, from the physical point of view the use of anticommuting 
variables yields some difficulties, that is why there was attempt to cure some of 
them, and modify the notion to the so called squbits \cite{castell} - 
prefix s comes here from abbreviated prefix super. 
Let us note that the term ``squbit'' is already used in literature in quite different context. 
In the nanotechnogy information technology it denotes superconducting 
qubit - a device based on Josephson's junction.

In the following we shall sketch the description of qubits in terms of the $\eta$-variables,
recall main properties of superqubits \cite{duff_1}. Then using results of 
Ref. \cite{castell} we will discuss briefly the modification 
of superqubit to the squbit. 

Before going into details let us comment on chronology of the main concepts
discussed in the present paper:
\begin{itemize}
\item 1976 - supersymmetric quantum mechanics introduced in 1976 by Nicolai \cite{nico}
\item 1970 - origins of quantum computing, qubits  
\item 2006 - qubits described by commuting nilpotent variables  \cite{mandi, amf-ncm}
\item 2009 - superqubits \cite{duff_1}
\item 2010 - squbits \cite{castell}
\end{itemize}
 
\section{Qubit - $\eta$-space description} 
The formalism of $\eta$-Hilbert spaces gives new tools to study multiqubit 
entanglement. Instead of considering the $\mathbb{C}^2$ Hilbert space and its tensor 
product structures one uses properties of functions of $\eta$-variables. 
Namely, if we take explicit form of one qubit and two qubit states in e.g. 
binary bases     
\begin{eqnarray}
\psi^{(1)}(x)&=&\psi_0(x)|0\rangle+\psi(x)|1\rangle,\\
\psi^{(2)}(x)&=&\psi_0(x)|00\rangle+\psi_1(x)|10\rangle+\psi_2(x)|01\rangle+\psi_{12}(x)|11\rangle ,
\end{eqnarray}
they can be written as the $\eta$-variable functions as follows 
\begin{equation}
\psi(x,\eta)=\psi_0(x)+\eta\psi(x)
\end{equation}
and similarly for two  qubits
\begin{equation}\label{2q}
\psi^{(2)}(x,\eta_1, \eta_2)=\psi_0(x)+\eta_1\psi_1(x)+\eta_2\psi_2(x)+\eta_1\eta_2\psi_{12}(x)
\end{equation}
In the case of pure states it provides strong tool - functional determinants 
and criteria for factorability of $\eta$-functions. This information is in direct correspondence
to the known measures of entanglement. 

The formalism can be briefly presented as follows. The $\eta$ variables are nilpotent elements coming from 
the commutative algebra $\mathcal{N}$ generated by unit and the set of nilpotents of  the first order. 
We can split given element to the numerical part and the rest   
$\mathcal{N}\ni\nu =b(\nu )+s(\nu)$ - the body and soul of an element. 
To describe the qubit and many-qubit states one takes 
bimodule over $\mathcal{N}$. $\mathcal{H}$ $\rightsquigarrow$ $\mathcal{N}$-module with the $\mathcal{N}$-scalar product i.e.
The generalized scalar product is defined as
\begin{equation}
<.\,,.>: \mathcal{H}\times \mathcal{H}\mapsto \mathcal{N}
\end{equation}
such that for $F,G\in \mathcal{H}$
\begin{eqnarray}
\nu^*< F,G>&=&<\nu F,G>\, =\, <F,\nu^* G>, \quad \nu\in \mathcal{N} \\
<F,G>&=&0 \quad\forall G\in \mathcal{H} \Rightarrow F=0\\
b(<F,G>)^*&=&b(<G,F>)\\
b(<F,F>)&\geq& 0, \quad\forall F \in \mathcal{H}
\end{eqnarray}
Particular realization of such $\mathcal{N}$-module is given by the function space $F[\vec{\eta}_n]$.
\begin{equation}
<F,~G>_{\mathcal{N}}=\int F^*(\vec{\eta})G(\vec{\eta})e^{<\vec{\eta}^*,\vec{\eta}>}~d\vec{\eta}^*~d\vec{\eta},
=\int F^*(\vec{\eta})G(\vec{\eta})d\mu(\vec{\eta}^*,~\vec{\eta})
\end{equation}
where
\begin{equation}
F^*(\vec{\eta})=\sum_{k=0}^n\sum_{I_k}F^*_{I_k}{\eta^{I_k}}^*,
\end{equation}
and $I_k=(i_1, i_2, \dots ,\i_k)$ is an ordered multi-index.
\begin{equation}
<F,~G>_{\mathcal{N}}=\sum_{k=0}\sum_{I_k}F^*_{I_k}G_{I_k}
\end{equation} 
Note that when $F_{I_k}\in \mathbb{C}$ the generalized $\mathcal{N}$-scalar product 
takes complex values and such assumption is natural for qubit's states description.  
\subsection{Entanglement of two qubits}
To illustrate how factorization criteria compare to entanglement let as take 
two qubit $\eta$-function (\ref{2q}). The $w_{12}$ denotes the  $\eta$-Wronskian 
with respect to $\eta_1$ and $\eta_2$ variables
\begin{equation}\label{w2eta}
w_{12}=det W_{12}=\left|
\begin{array}{cc}
F&\frac{\partial F}{\partial \eta_1}\\
\frac{\partial F}{\partial \eta_2}&\frac{\partial^2 F}{\partial \eta_1\partial \eta_2}
\end{array}
\right|=
\left|
\begin{array}{cc}
F&\partial_1 F\\
\partial_2 F&\partial_{12} F
\end{array}
\right|=F_0F_{12}-F_1F_2
\end{equation}
There is valid the following statement: for arbitrary function  $F(\eta_1, \eta_2)$ 
vanishing of the Wronskian $w_{12}(F)$ is equivallent to factorization of the form 
$F(\eta_1, \eta_2)=G(\eta_1)\tilde{G}(\eta_2)$, for some $G$ and
$\tilde{G}$. Wronskian for the function of two $\eta$ variables
has numerical values. 
Considering the Werner state represented by the function
$\psi_W(\eta_1, \eta_2)=\frac{1}{\sqrt{2}}(\eta_1+\eta_2)$ we get that  $w_{12}(\psi_W)=-\frac{1}{2}$.
For the GHZ state $\psi_{GHZ}(\eta_1, \eta_2)=\frac{1}{\sqrt{2}}(1+\eta_1\eta_2)$ and
$w_{12}(\psi_{GHZ})=\frac{1}{2}$

The two-tangle expressed by the Wronskian takes the form
\begin{equation}
\tau(\psi)=4|w_{12}(\psi)|,
\end{equation} 
e.g. $\tau(\psi_W)=1$, $\tau(\psi_{GHZ})=1$.
This approach works for multi-qubit states. For three qubits and more 
the Wronskians depend explicitly on $\eta$-variables, what yelds the 
array of conditions and selection of important invariants for multi-qubit 
states. What is intersting the constraints comming from these criteria are
desirable, because they select a subfamily of important invariants from the
very quickly growing with the number of qubits, the set of all invariants.
Detailed account of this approach can be found 
in the Refs. \cite{amf-fe, amf-nqm}.

\section{Superqubit - extension of qubit to $(2|1)$ superspace}

The superspace formalism turns out to fit the description of the extension of qubit
to the system which is symmetric with respect of the $Osp(1|2)$, the minimal 
supersymmetric extension of the $SL(2, \mathbb{C})$ which is the group SLOCC 
transformations playing the fundamental role in description of entanglement.
This was the departure point in the definition of superqubit given in Ref. \cite{duff_1}.
Let the $\mathcal{Q}=\mathcal{Q}_0+\mathcal{Q}_1$ denotes the Banach-Grassmann algebra
in the sense of Rogers, where $\mathcal{Q}_i$ is its even and odd part for $i=0$ 
and $i=1$ respectively. The numerical part of $q\in\mathcal{Q}$ is called body the 
rest is a soul of the element $q$: $q =b(q )+s(q)$. One considers a graded module 
$\mathbb{V}=\mathbb{V}_0\oplus\mathbb{V}_1$ of dimension $(2|1)$ over the algebra
$\mathcal{Q}$ which extends the $\mathbb{C}^2$ Hilbert space of a qubit.
\begin{equation}
\mathbb{C}^2 \rightsquigarrow \mathbb{V}^{(2|1)}_{\mathcal{Q}} 
\end{equation}
with the $\mathcal{Q}$-scalar product i.e.
\begin{equation}
<.\,,.>: \mathbb{V}\times \mathbb{V}\mapsto \mathcal{Q}
\end{equation}
such that for $\psi$,$\phi \in \mathbb{V}$
\begin{eqnarray}
< \psi,\phi q>&=&<\psi,\phi>q\, \quad q\in \mathcal{Q} \\
<\psi,\phi>^{\#}&=&(-1)^{\psi(\phi+1)}<\phi,\psi>,
\end{eqnarray}
where $(q^{\#})^{\#}=(-)^{|q|} q$ for homogenous elements 
i.e. $q\in\mathcal{Q}_i$, $|q|=i$, $i=0,1$.

The one-superqubit system is described by supervector
\begin{equation}
\psi=|0\rangle \psi_0+|1\rangle \psi_1+|\bullet\rangle \psi_{\bullet}=
|\psi\rangle=|k\rangle\psi_k+|\bullet \rangle \psi_{\bullet}
\equiv \sum_X |X\rangle \psi_X
\end{equation}  
where $|i\rangle$, $i=0,1$ are even vectors, $\psi_i\in\mathcal{Q}_0$ and   
$|\bullet\rangle$ is an odd vector, $\psi_{\bullet}\in\mathcal{Q}_1$. In this way 
the two level system is extended to the three level system with one level being of different nature
then the other two. Moreover, there is no balance between
odd and even (bose and fermi) degrees of freedom as in usual supersymmetric models. 
The $\mathcal{Q}$-scalar square  
$\langle\psi,\psi\rangle=\delta^{ij} \psi_i^{\#}\psi_j-\psi_{\bullet}^{\#}\psi_{\bullet}$
gives value with the nontrivial soul and normalized superstate has the form
\begin{equation}
|\psi\rangle=(\delta^{i_1i_2}\psi^*_{i_1}\psi_{i_2})^{-\frac{1}{2}}
\left(|k\rangle (1+\frac{1}{2}(\delta^{i_1i_2}\psi^*_{i_1}\psi_{i_2})^{-1}
\psi_{\bullet}^{\#}\psi_{\bullet})\psi_k
+|\bullet \rangle \psi_{\bullet}\right)
\end{equation}
\subsection{Two superqubits}
To see how questions concerning entanglement can be studied let us consider 
a two-superqubit states \cite{duff_1}
\begin{equation}
|\psi\rangle=|jk\rangle\psi_{jk}+|j\bullet\rangle\psi_{j\bullet}+
|\bullet k\rangle\psi_{\bullet k}+|\bullet\bullet\rangle \psi_{\bullet\bullet}
\end{equation} 
$dim \mathbb{V}^{(2)}=(5|4)$;  $\psi_jk$, $\psi_{\bullet\bullet}$ are even 
and $\psi_{\bullet k}$, $\psi_{j\bullet}$ are odd. The super-SLOCC now is given 
by $OSp(2|1)\times OSp(2|1)$. To define the generalization of the entanglement
for two-superqubits we can use the $(2|1)\times (2|1)$-supermatrix
\begin{equation}
\psi_{XY}=\left(
\begin{array}{cc}
\psi_{jk} & \psi_{j\bullet}\\
\psi_{\bullet k} & \psi_{\bullet\bullet}
\end{array}
\right).
\end{equation}
Natural generalization of determinant is the Berezinian, but it is not good candidate 
for super-generalization of 2-tangle. Its explicit form
\begin{equation}
Ber (\psi_{XY})= det(\psi_{jk}-\psi_{j\bullet}\psi^{-1}_{\bullet\bullet}
\psi_{\bullet k})\psi^{-1}_{\bullet\bullet}
\end{equation}
shows that invertibility of $\psi_{\bullet\bullet}$ is too strong condition, restricting 
the possible two-superqubit states, since for example for the product state 
the $\psi_{\bullet\bullet}$ will be a product of two odd Grassmann numbers and therefore noninvertible. 
The solution given in Ref. \cite{duff_1} is based on the following observation done 
for conventional determinant    
\begin{equation}
det(\psi_{jk})=\frac{1}{2}tr((\psi\epsilon)^T\epsilon \psi)
\end{equation}
then using the following modifications
($SL(2)$ to $Osp(2|1)$) $\oplus$ (transpose $\rightarrow$ super-transpose)
one can write
\begin{equation}
sdet (\psi_{XY})=\frac{1}{2}str((\psi E)^{ST}E \psi),
\end{equation}
what gives for two-superqubits explicitly
\begin{equation}
sdet (\psi_{XY})= (\psi_{00}\psi_{11}-\psi_{01}\psi_{10}+\psi_{0\bullet}\psi_{1\bullet}
+\psi_{\bullet0}\psi_{\bullet 1})-\frac{1}{2}\psi_{\bullet\bullet}^2
\end{equation}
This expression turned out to be good candidate for generalization of 2-tangle 
to the super 2-tangle 
\begin{equation}
s\tau_{XY}=4sdet \psi_{XY}(sdet \psi_{XY})^{\#}
\end{equation}
It has basics property that $s\tau_{XY}=0$ for product states. However its behaviour
is different then we have used to for 2-tangle. Let us illustrate this using some examples:
\begin{itemize}
\item[\em Ex.1.] The simple superstate \cite{duff_1}
$$
|\phi\rangle=i|\bullet\bullet\rangle, 
$$
is maximally entangled in the sense of super 2-tangle: $s\tau_{XY}=1$.
This is an "algebraic artefact" because the separation of a superstate
proportional to $|\bullet\bullet\rangle$ to product of one-superqubit states
requires the nilpotent (a bodyless) coefficient in front of it being the product of two 
odd Grassmann numbers. Let us call such a state algebraically entangled.
\item[\em Ex.2.] Maximally super-entangled superstate \cite{duff_1}  
\begin{equation}
|\psi\rangle=\frac{1}{\sqrt{3}}(|00\rangle+|11\rangle+i|\bullet\bullet\rangle)
\end{equation}
$$
s\tau_{XY}(|\psi\rangle)=4(\frac{1}{3}+\frac{1}{2}\frac{1}{3})^2=1,
$$
\item[\em Ex.3.] similar superstate
\begin{equation}
|\psi\rangle=\frac{1}{\sqrt{3}}(|00\rangle+|11\rangle+|\bullet\bullet\rangle)
\end{equation}
is not maximally entangled and $s\tau_{XY}=\frac{1}{9}$. 
\item[\em Ex.4.] The following superstate 
\begin{equation}
|\psi\rangle=\frac{1}{2}(|00\rangle+|11\rangle+\sqrt{2}|\bullet\bullet\rangle)=
\frac{1}{\sqrt{2}}(\psi_{GHZ}+|\bullet\bullet\rangle)
\end{equation}
has vanishing super 2-tangle, while it seems to be not separable 
($\psi_{GHZ}$ is the two qubit Greenberger – Horne – Zeilinger state). 
\end{itemize}
%
Let us note that there exists generalization of the above construction to the 
three superqubits: $s\tau_{XYZ}$ given in the Ref. \cite{duff_1}. 

This approach is 
under development, recently there were proposed two-superqubit states that are 
"more nonlocal" then respective two-qubit states \cite{duff_2}. The analysis involves operation
with the Grassmann valued probabilities and mentioned before choice of the way of ascribing
final numerical values to such an objects - that always was a problematic issue 
of SUSY-QM in superHilbert spaces.  
%
\section{Squbit - reshaped superqubit}
Soon after the introduction of the superqubit there was proposed its modification \cite{castell},
with aims to cure the drawback of superqubits:  the Grassmannian coordinates make difficult/impossible 
a probabilistic interpretation. 
The proposed modification to ged rid of Grassmann valued probability amplitudes requires
the introduction of auxiliary Grassmann variables $\theta_i$ in such a way that new superstate can be written
in the form 
\begin{equation}
|\psi \rangle = b | B \rangle + f_i \theta_i | F_i \rangle +
b_{ij} \theta_i \theta_j | B_{ij} \rangle + f_{ijk} \theta_i
\theta_j \theta_k  | F_{ijk} \rangle + \cdots
\end{equation}
$b$, $f$ are complex; $|B\rangle$ are even; $|F\rangle$ are odd.
Such  $\theta$ variables "neutralize" odd states $|F\rangle$.
Then
\begin{eqnarray}
\langle \langle \psi | \psi \rangle \rangle &\equiv&  \int e^{\sum_i \bar{\theta}_i \theta_i} \langle
\psi | \psi \rangle~
\Pi_i d\bar{\theta}_i d\theta_i =\\ \nonumber
&&|b|^2 + \sum |f_i|^2  +  \sum_{i<j} |b_{ij}|^2 + \sum_{i<j<k}|f_{ijk}|^2  + \cdots\nonumber
 \end{eqnarray}
restores possibility of probabilistic interpretation. Let us note that it resembles
earlier studied approach based on the commuting nilpotent $\eta$-variables.

Now, the superspace of superstates has the decomposition
${\cal H}_{BF} = {\cal H}_B \oplus {\cal H}_F$
\begin{itemize}
\item bosonic states: $|0\rangle, \theta_i \theta_j |0\rangle, \dots$ 
\item fermionic states: $\theta_i |0\rangle, \theta_i \theta_j \theta_k|0\rangle,..$, 
\end{itemize}
where $|0\rangle$ is the Fock vacuum.
Moreover one complements set of $\theta_i$ by $\bar{\theta}^j$ in such a way that
$\theta_i $, $\bar{\theta}^j$ satisfy the Clifford algebra
\begin{equation} 
\{ \bar\theta^{i}, \theta_j\} =\delta^i_{j},\quad
\{ \theta_{i}, \theta_j\} = 0,\quad \{ \bar\theta^{i}, \bar\theta^j\} = 0
\end{equation}
and the new space ${\cal H}^\theta={\cal H}^\theta_B \oplus {\cal H}^\theta_F$ 
is spanned on 
\begin{equation}
\theta_i \theta_j \cdots |0 \rangle \otimes \theta_k \theta_l \cdots |0 \rangle =
\theta_i \theta_j \cdots  \theta_k \theta_l \cdots |0 \rangle \label{tensortheta}
\end{equation}
To conlude the construction the extension of  ${\cal H}_{BF}$ 
to ${\cal H}_{BF} \otimes {\cal H}^\theta$ is proposed
\begin{equation}
{\cal H} = \Big({\cal H}_B \otimes
{\cal H}^\theta_B\Big) \oplus \Big({\cal H}_F \otimes {\cal
H}^\theta_F\Big)
\end{equation}
giving the bosonic and fermionic dimensions of this space balanced i.e.
\begin{itemize}
\item  $2^{N-1}$ of $|B_I\rangle$
\item  $2^{N-1}$ of $|F_I\rangle$
\end{itemize}
Finally superstate defining the squbit can be written in the following form
\begin{equation}
|\psi\rangle = b|0\rangle \otimes |B\rangle + f_i \theta_i
|0\rangle \otimes |F_i \rangle + b_{ij} \theta_i \theta_j
|0\rangle \otimes |B_{[ij]} \rangle + f_{ijk} \theta_i \theta_j
\theta_k |0\rangle \otimes |F_{[ijk]} \rangle + \dots \label{cliff}
\end{equation}
What gives the simplest squbit:
\begin{equation}
| \psi\rangle =  b |0\rangle \otimes |B\rangle + f \theta
|0\rangle \otimes |F \rangle
\end{equation}
Let us bring here the conclusion of the Authors of the Ref. \cite{castell}: {\em  
the auxiliary $\theta$ system allows the dynamical mixing between fermions and bosons of the
${\cal H}_{BF}$ system, with a well-defined positive norm, necessary for a probabilistic 
interpretation}. 

As it is seen from this brief account, the squbit has ballanced fermionic and 
bosonic degrees of freedom and is more close to usual supersymmetric models. 
It also seems
to departure too far from the original superqubit model. 
\section*{Final Comments}
The notion of qubit is deeply rooted in quantum mechanics and the importance 
of this simple system was discovered after realizing the existence of new resource
available in nature - quantum entanglement. For decades the real value of quantum 
entanglement was neglected and occasionaly studied by people interested in quantum 
paradoxes and fundamental problems of the quantum mechanics. Presently it is 
intesively studied theoretically and its evidence is experimentally determined. 
Despite research based on conventional for quantum mechanics formalism of 
the Hilbert space, there are atempts to describe entanglement in different ways,
using experiences from supermathematics and other algebraical structures.
The $\eta$-Hilbert space formulation based on commuting nilpotent variables 
gives new functional tools to study many-qubit entanglement and to realize 
that many of interesting entangled states can be identified as elementary 
$\eta$-functions \cite{amf-fe, amf-trig}. On the other hand after four decades of supersymmetric theories
there was proposed super-extension of the notion of qubit and super-entanglement
of superstates with theoretically more nonlocality then in conventional 
many-qubit states. But physical meaning of this extension is still unclear. 
%

\end{document}